\documentclass[superscriptaddress, showpacs, floatfix ,reprint]{revtex4-1}
\usepackage{amssymb}
\usepackage{amsmath}
\usepackage{times}
\usepackage{hyperref}
\usepackage{graphicx}
\usepackage{epsfig}
\usepackage{float}

\def\ie{\mbox{\em i.e.\ }}
\def\eg{\mbox{\em e.g.\ }}
\def\uu{\mathbf u}
\def\bb{\mathbf b}
\def\ff{\mathbf f}
\def\xx{\mathbf x}
\def\kk{\mathbf k}
\def\pp{\mathbf p}
\def\qq{\mathbf q}
\def\si{\mathcal{\scriptstyle{K}}}          
\def\sii{\mathcal{\scriptstyle{P}}}        
\def\siii{\mathcal{\scriptstyle{Q}}}    
\def\kp{p}   
\def\kc{c}

\begin{document}

\title{On the locality of MHD turbulence scale fluxes}
\author{Bogdan Teaca}
\email{bogdan.teaca@epfl.ch, Tel: +41-21-693.43.05}
\affiliation{Centre de Recherches en Physique des Plasmas, Science de Base, Ecole Polytechnique Federale de Lausanne, Station 13, Building PPB
CH-1015 Lausanne, Switzerland.}
\author{Daniele Carati}
\email{dcarati@ulb.ac.be, Tel: +32-2-650.58.13}
\affiliation{Statistical and Plasma Physics, Faculty of Sciences, Universit\'e Libre de Bruxelles, Campus Plaine, CP 231, B-1050 Brussels, Belgium.}
\author{J. Andrzej Domaradzki}
\email{jad@usc.edu, Tel: 213-740-5357}
\affiliation{Department of Aerospace and Mechanical Engineering, University of Southern California, Los Angeles, CA 90089-1191, USA.}
\begin{abstract}
The scale locality of energy fluxes for magnetohydrodynamics (MHD) is investigated numerically for stationary states of turbulence. Two types of forces are used to drive turbulence, a kinetic force that acts only on the velocity field and a kinetic-inductive forcing mechanism, which acts on the velocity and magnetic fields alike. The analysis is performed in spectral space, which is decomposed into a series of shells following a power law for the boundaries. The triadic transfers occurring among these shells are computed and the fluxes and locality functions are recovered by partial summation over the relevant shells. Employing Kraichnan locality functions, values of $1/3$ and $2/3$ for the scaling exponents of the four MHD energy fluxes are found. These values are smaller compared with the value of $4/3$ found for hydrodynamic turbulence. To better understand these results, an in depth analysis is performed on the total energy flux. 
\end{abstract}
\pacs{47.65.-d, 47.27.ek, 47.27.er} 
\maketitle

\section{Introduction}

Energy transfer functions in turbulence are the result of nonlinear interactions among different scales of motion. Although all the scales of the flow are coupled together in the same manner, through the nonlinear terms, the global transfer of energy to a scale is dominated by contributions from particular scales. Placing adequate bounds on the scales that bring the most contributions to the energy flux through a scale is crucial for the developing of adequate turbulence models, like Large Eddy Simulations (LES), Shell Models and for advancing physical understanding of turbulent phenomenology. Finding these limits is the main object of locality analysis for strongly coupled nonlinear systems. 

In general, in an effort to understand the behavior of scale coupling for any system, the resulting transfers due to the nonlinear terms are investigated for global conserved quantities. The redistribution nature of such conserved quantities in spectral space provides insights into the nonlinear dynamics of the system. In the absence of such global conserved quantities, as in the case of dissipative systems, ideal invariant quantities are used instead, \ie quantities that become conserved when the dissipative terms are taken to be exactly zero. For dissipative systems, an external force that acts as a source for the ideal invariant quantity is employed and statistically stationary states of the system in regard to the ideal invariant quantity can be achieved. 

For hydrodynamic (HD) turbulence, the energy represents such an ideal invariant quantity along with kinematic helicity. In the inertial range, defined as the interval of scales smaller than that on which external forces act on and larger than scales where dissipative effects dominate, the properties of energy transfer functions, which depend exclusively on the nonlinear dynamics of turbulence, are expected to have a general, universal behavior. For a plasma medium, the nonlinear self-coupling problem of large spatio-temporal scales can be investigated in the magnetohydrodynamic (MHD) limit. For MHD turbulence, three quadratic ideal invariants exist: the total energy, cross-helicity and magnetic-helicity. Although strictly speaking only the total energy is an ideal invariant for MHD, it is often interesting to know the behavior of both the kinetic and magnetic energy transfer channels.

In MHD turbulence, the locality problem is more complex. Not only multiple energy fluxes exist compared to just one for HD turbulence, but different dynamical states are now possible, providing different behaviors for the fluxes across scales. Traditionally, with each scale of motion $\ell$, which can denote the size of an eddy, a wavenumber $k\sim1/\ell$ is associated and the interactions between fields filtered to correspond to different scales $\uu_k(\xx,t)$ and $\bb_k(\xx,t)$ are investigated. Alternatively, the same ideas can be applied to the Els\"asser representation\cite{Elsasser:1950p424} of the fields for which the nonlinear terms can be interpreted physically as the scattering of contra-propagating Alfv\'en waves. In this approach, a Fourier space decomposition of the velocity field becomes the natural framework and the transfer function between scales are obtained by selective integration (filtering) over the Fourier modes denoted by the wavevectors $\kk$. A review on the works employing this approach for MHD turbulence was done recently by Mininni\cite{Mininni:2010p1106}.

Another way of looking at the problem is to relate the locality properties of the energy fluxes to the scaling of the velocity and magnetic fields.
It is common\cite{Zhou:2004p21,Zhou:2005p976} to consider the order of the fluctuation for a scale $\delta v_k$ to depend on the energy $E(k)$ as $\delta v_k=[kE(k)]^{1/2}$. Assuming a Kolmogorov type scaling for the energy spectrum in the inertial range, $E_K(k)=C_K \varepsilon^{2/3}k^{-5/3}$, we obtain the scaling for the velocity field to be $\delta v_k \sim k^{-1/3}$. Using the Iroshnikov-Kraichnan form of the energy spectrum characteristic of weakly Alfv\'enic  turbulence, $\displaystyle{E_{IK}(k)=C_{IK}(V_A \varepsilon)^{1/2}k^{-3/2}}$, we find $\delta v_k \sim k^{-1/4}$, where $V_A=B/\sqrt{4\pi \rho}$ is the Alfv\`{e}n speed of the guide magnetic field. Since these two scaling laws correspond to the two known limits of MHD turbulence\cite{Galtier:2005p92}, we can write in general that $\delta v_k\sim k^{-\varsigma}$, where $\varsigma\in[1/4, 1/3]$. 

Using the smoothness condition for the MHD fields, which can be seen as arising from the scaling index $\varsigma $, Aluie and Eyink showed\cite{Aluie:2010p946} that for a dyadic (octave) separation of scales, the energy fluxes in MHD are local and can be characterized in the infrared (IR) and ultraviolet (UV) ranges by the scaling exponent with a limit depending on $\varsigma$ of the two fields. In particular for $\varsigma=1/3$ the IR and UV limits are found to be $\pm 2/3$. Since these limits are related to the MHD scaling, they are quite robust as results. 

In a work\cite{Domaradzki:2010p1012}, the present authors reported a locality exponent value of $1/3$ for the energy conversion flux appearing in the kinetic energy equation and which that is responsible for the conversion of kinetic energy into magnetic energy. This limit would imply a stronger nonlocal effect for MHD turbulence. Since in the past it was suggested that this value could be a result of the external force polluting the inertial-inductive range, additional simulations have been performed and the values of the locality exponents are readdressed in this paper.

\section{MHD stationary states}

The locality properties of the scale fluxes are investigated for stationary state solutions of the incompressible MHD equations:
\begin{align}
&\frac{\partial \uu}{\partial t} = - \uu \cdot \nabla \uu + \bb \cdot \nabla \bb + \nu \nabla^2 \uu + \ff^u  - \nabla p\;, \label{vel}\\
&\frac{\partial \bb}{\partial t} = - \uu \cdot \nabla \bb + \bb \cdot \nabla \uu  + \eta \nabla^2 \bb +\ff^b \;, \label{mag} \\
&\nabla \cdot \uu=0\; ,\ \nabla \cdot \bb=0 \;, \label{divfree}
\end{align}
where $\uu=\uu(\xx,t)$ is the fluid velocity field, $\bb=\bb(\xx,t)$ is the magnetic field expressed in Alfv\`{e}n units and $p=p(\xx,t)$ is the total (hydrodynamic + magnetic) pressure field divided by the constant mass density, $\rho$. Due to the incompressibility condition, the pressure $p$ is an auxiliary variable and can be formally eliminated by solving the Poisson equation,
\begin{align}
\nabla^2 p=-\nabla \uu : \nabla \uu+\nabla \bb : \nabla \bb \;.
\end{align}
Throughout this work, the fluid viscosity $\nu$ and the magnetic diffusivity $\eta$ are taken to be equal. 

The zero divergent, external force fields $\ff^u=\ff^u(\xx,t)$ and $\ff^b=\ff^b(\xx,t)$ act on the velocity and magnetic fields, respectively. The two forces are part of a forcing mechanism that imposes the injection rates of the MHD ideal invariant quantities. We will refer to this type of forcing mechanism as a {\em kinetic-inductive} force. A kinetic only forcing method ($\ff^u \equiv \ff$ and $\ff^b \equiv 0$), used previously in the literature for similar studies \cite{ Carati:2006p632, Teaca:2009p628}, is also employed. In Fourier space, the forces $\hat{\ff}^u(\kk)$ and $\hat{\ff}^b(\kk)$ are assumed to be local, zero divergent quantities that act equally on all the modes within the wavenumber shell $s_{f}=[k_{\inf }, k_{\sup}]$. Usually, the shell $s_{f}$ is considered at large scales and is sufficiently thick to contain a large number of modes so that no preferential direction is introduced in the flow. In essence, both the kinetic only force $\ff$ and the kinetic-inductive forcing mechanism $\ff^{\{u,b\}}$ are considered to be proportional to the fields as, 
\begin{align}
\hat{\ff}(\kk) &=C_1(\kk) \hat{\uu}(\kk) + C_2(\kk) \hat{ \boldsymbol \omega}(\kk) \;,
\end{align}
where $\boldsymbol{\omega}=\nabla \times \uu$ is the vorticity and
\begin{align}
\hat{\ff}^{\{u,b\}}(\kk) &=C_1^{\{u,b\}}(\kk) \hat{\uu}(\kk) + C_2^{\{u,b\}}(\kk) \hat{\bb}(\kk) \;.
\end{align}
%
\begin{figure}[b]
\centering
\includegraphics[width = 0.482\textwidth]{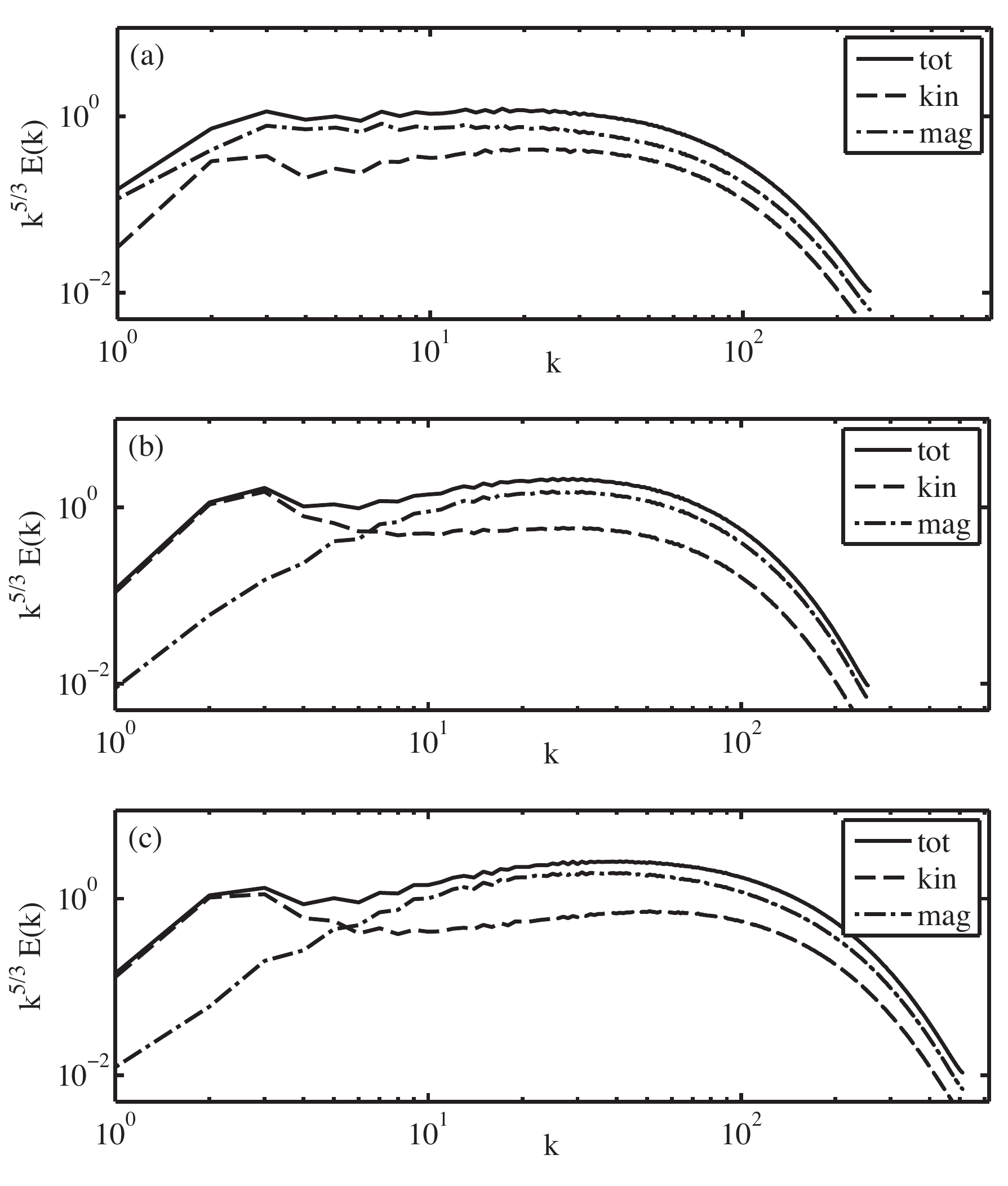}
\caption{The compensated energy spectra for run I (a), run III (b) and run IV (c).}
\label{fig_spec}
\end{figure}
%
\begin{table*}[tb]
\begin{tabular}{c| c c c c c c c | c c c c c }
\hline
\hline
 \ \ \ Run \ \ \ & \ \ \  Nr. of modes  \ \ \ & \ \ \  $k_{\max}$  \ \ \ & \ \ \   $\nu=\eta$ \ \ \ & \ \ \  Force Type \ \ \ &  \ \ \ $[k_{inf}, k_{sup}]$  \ \ \ &  \ \ \  $\varepsilon$  \ \ \ &  \ \ \  $\sigma$  \ \ \  & \ \ \   $R_\lambda$ \ \ \ &  \ \ \ $r_A$  \ \ \ &  \ \ \ $L^u$  \ \ \ &  \ \ \ $L^b$ \ \ \ &  \ \ \ $k_{\max}\eta$      \ \ \ \\
\hline
I    & $512^3$   &  256   &  0.00050 & $\ff^u \&\ \ff^b$ & [2.5, 3.5] &$0.1$  & $0.0$&$217$  &$0.47$  &$1.55$  &$1.86$   &$1.05$      \\
II   & $512^3$   &  256   &  0.00050 & $\ff^u \&\ \ff^b$ & [2.5, 3.5] &$0.1$  & $0.4$&$296$  &$0.58$  &$1.92$  &$1.94$   &$1.06$      \\
III  & $512^3$   &  256   &  0.00055 & $\ff^u$ only       & [1.5, 3.1] &$0.32$ & $0.0$&$362$ &$1.96$  &$1.76$  &$0.37$   &$1.22$     \\
IV  & $1024^3$ &  512   &  0.00030 & $\ff^u$ only       & [1.5, 3.1] &$0.32$ & $0.0$&$450$ &$1.72$  &$1.94$  &$0.35$   &$1.15$     \\
\hline
\hline
\end{tabular}
\caption{The simulations control parameters and the relevant diagnostics at the moment the flux analysis is being performed. The diagnostics are defined as: $R_\lambda=2\pi u^{\mbox{\scriptsize rms}} \sqrt{\frac{\int dk E^u(k)}{\nu^2 \int dk k^2 E^u(k)}}$; $r_A=E^u/E^b$; $L^{\{u,b\}}=2\pi \frac{\int dk k^{-1} E^{\{u,b\}}(k)}{ \int dk E^{\{u,b\}}(k)}$ and $\eta=[\nu^3/\varepsilon]^{1/4}$.}
\label{database} 
\end{table*}
The real valued parameters $C_1$ and $C_2$ are obtained by requiring the force to inject into the system a certain amount of energy in unit time (power) $\varepsilon$ and zero kinetic helicity. Similarly, for the kinetic-inductive version of the force the $C_1^{\{u,b\}}$ and $C_2^{\{u,b\}}$ parameters are obtained by selecting the amount of energy and cross-helicity ($\varepsilon \sigma$) injected into each of the two fields. The cross-helicity parameter $\sigma$ is bounded in the interval $[-1,1]$. Since the forces parameters are real, the forcing methods do not influence phases of the fields, which ensures that no change is made in the type of turbulent structures present in the system. This might generate different results compared to the case of injecting cross-helicity by imposing the alignment of $\uu$ and $\bb$ in the real space for example. The advantage of the type of forces used in our work is in the ability to select the levels at which the ideal invariant fluxes will relax to once the stationary state is achieved. Since the forces are proportional to the fields, the characteristic time of the force will tend to be equal to that of the nonlinear cascade for a scale located in $s_f$, independent of the energy injection level selected. A more detailed description of the electromagnetic forcing mechanism is given in another work\cite{Teaca:2011p1338}.

The equations (\ref{vel}-\ref{mag}), with the appropriate choice of forcing, are solved using a pseudo-spectral solver, using periodic boundaries conditions in all three directions. The size of the domain in each direction is $2\pi$. The solver uses a FFT algorithm for the space discretisation and a order 3rd Williamson-Runge-Kuta method for the time advancement. The time step is computed automatically to be consistent with the CFL criterion. The nonlinear terms are partially dealiased using a phase-shift method~{\cite{Patterson:1971p965}. The simulations are run until a statistically stationary state is reached for both the velocity and the magnetic field. Multiple runs were made, details being listed in Table~\ref{database} for the well resolved stationary states investigated. Run IV is obtained from run III by increasing the numerical resolution and decreasing the viscosity and magnetic diffusively accordingly. The compensated spectra for the runs considered are displayed in Fig.\ref{fig_spec}. Although all the data has been investigated, runs I and III will be mostly used for data display to exemplify best the effects generated by the two type of forces.

\section{Locality framework}

\subsection{Energy equation for a mode}

Since we are interested in the study of locality between scales, the MHD equations are solved in Fourier space. The energy equations for a mode $\kk$ are easily derived and read as:
\begin{align}
&\frac{\partial E^u(\kk)}{\partial t} = T_{u,u}^u(\kk) - T_{b,b}^u(\kk) - 2\nu k^2 E^u(\kk)   + I^u(\kk), \label{Evel}\\
&\frac{\partial E^b(\kk)}{\partial t} = T_{b,u}^b(\kk) - T_{u,b}^b(\kk)  - 2\eta k^2 E^b(\kk) + I^b(\kk)\;, \label{Emag}
\end{align}
Each of the non-linear terms of the type $-\bf{Z} \cdot \nabla \bf{Y}$ appearing in the right hand side of the field $\bf{X}$ evolution equations, will generate energy transfers in spectral space of the form:
\begin{align}
T^X_{Y,Z}(\kk)=\iint d\qq\ d\pp\ \Re &\left\{ i\kk \cdot \hat{ \bf Z}(\qq) \hat{\bf Y}(\pp)\cdot \hat{\bf X}(\kk) \right\}\times \nonumber\\
& \delta(\kk+\pp+\qq)\;, \label{transfer}
\end{align}
where $\Re$ stands for the real part of a complex number and we have used the reality condition $\hat{\bf X}^*(\kk)=\hat{\bf X}(-\kk)$ which results in $T^X_{Y,Z}(\kk)=T^X_{Y,Z}(-\kk)$ for the real valued transfer. The field notations $\{X,Y,Z\}$ stand in for $u$ or $b$ depending on the specific transfer and only their position in the transfer expression is important. While $Z$ represents the advecting fields and $Y$ the advected field, $X$ stands in for the receiving field. The delta Dirac function limits the transfers to wave-vector triads that satisfy the conditions $\kk+\pp+\qq=0$. Note that in our previous work on this topic\cite{Domaradzki:2010p1012}, terms in the magnetic energy equation were denoted by $T_{ub}$ and $T_{bu}$, i.e., with subscripts $u$ and $b$ interchanged compared with notation used in (\ref{transfer}). 

The equations (\ref{Evel}-\ref{Emag}) also contain energy injection terms $I^X(\kk)$ which are due to the external forces and have the form:
\begin{align}
I^X(\kk)&=\Re \left\{\ff^X(\kk) \cdot \hat{\bf X}(-\kk)\right\} \;, \label{injection}
\end{align}
For stationary state turbulence, the total energy injection level imposes the total flux level in the inertial range and is given by the summation of the two contributions: $I^u+I^b$. Individually, $I^u$ and $I^b$ do not constrain respectively the the kinetic and magnetic energy flux levels, as the kinetic and magnetic energy are not themselves conserved quantities. For stationary state turbulence, the kinetic and magnetic energy fluxes relax to a level constrained by the kinetic and magnetic dissipation rates, respectively.

\subsection{Triad transfers and conservation properties}

From Eq. (\ref{transfer}), we see that the net energy received by the mode $\kk$ of field $X$ is due to the interaction with all possible modes $\pp$ and $\qq$ which form a triad. It is useful to look at the energy transfer for a individual triad, defined as: 
\begin{align}
T^X_{Y,Z}(\kk|\pp,\qq)=\frac{1}{2}\Re\{i\kk \cdot [\hat{ \bf Z}(\qq)\hat{\bf Y}(\pp)+\hat{ \bf Z}(\pp)\hat{\bf Y}(\qq)] \cdot \hat{\bf X}(\kk)\}\;, \label{triad_tra}
\end{align}
for $\kk+\pp+\qq=0$ and zero otherwise. Since the triad transfer function is symmetric in $\pp$ and $\qq$, \ie $T^X_{Y,Z}(\kk|\pp,\qq)=T^X_{Y,Z}(\kk|\qq,\pp)$, we have expressed this in an explicit way in the definition (\ref{triad_tra}). Because of this symmetry, although we know the role played by each field in the interaction we have no way of differentiating between the contribution of modes $\pp$ and $\qq$ in a unique way. Following the work of Dar and Verma \cite{Dar:2001p209, Verma:2004p206}, a mode-to-mode transfer function can be introduced by accounting for a circulating transfer (an uncertainty) that cancel itself exactly for the triad transfer,  
\begin{align}
S^X_{Y,Z}(\kk|\pp|\qq)=\Re\{ [i\kk \cdot &\hat{ \bf Z}(\qq)][ \hat{\bf Y}(\pp)\cdot \hat{\bf X}(\kk)] \}\;,\label{mode_tra}
\end{align}
 for $\kk+\pp+\qq=0$ and zero otherwise. The triad transfer can be expressed now as the sum of two mode-to-mode transfers:
\begin{align}
T^X_{Y,Z}(\kk|\pp,\qq)&=\frac{1}{2}\left\{ S^X_{Y,Z}(\kk|\pp|\qq)+S^X_{Y,Z}(\kk|\qq|\pp) \right\} \;. 
\end{align}
For the energy transfer, although the nonlinear terms have the same form, they have different physical significance. The terms where $u$ is the advective quantity, are respectively responsible for the kinetic and magnetic energy conservation. This fact is expressed using the triad transfers as a conservation of interaction between the modes that make up a triad:
\begin{align}
T^u_{u,u}(\kk|\pp,\qq)+T^u_{u,u}(\pp|\qq,\kk)+T^u_{u,u}(\qq|\pp,\kk)=0 \;, 
\end{align}
and
\begin{align}
T^b_{b,u}(\kk|\pp,\qq)+T^b_{b,u}(\pp|\qq,\kk)+T^b_{b,u}(\qq|\pp,\kk)=0 \;. 
\end{align}
The other two terms, where $b$ is the advective quantity, do not conserve the energy individually as they represent transfers from one field to the other ({\em cross-field transfers}). However, their sum, which accounts for the cross-field transfer, is conserved for any triad,
\begin{align}
&T^u_{b,b}(\kk|\pp,\qq)+T^u_{b,b}(\pp|\qq,\kk)+T^u_{b,b}(\qq|\pp,\kk)+ \nonumber \\
&T^b_{u,b}(\kk|\pp,\qq)+T^b_{u,b}(\pp|\qq,\kk)+T^b_{u,b}(\qq|\pp,\kk)=0 \;.
\end{align}
The cross-field transfers are responsible for the conversion of kinetic energy into magnetic one and vice-versa and it is due to their existence that the kinetic and magnetic energies are not conserved individually. This aspect needs to be considered when looking at the fluxes generated by the two terms taken separately. 

As a side note, we see that the same mechanism exists for cross-helicity, which has units of energy and up to a point can be considered as the energy contained by the velocity magnetic interaction. For the cross-helicity, the transfers $T^b_{b,b}$, $T^u_{u,b}$  account for the transfer of information for the same field, while $T^b_{u,u}+T^u_{b,u}$ represents the transfer between the velocity and magnetic field. These cross-helicity interactions are conserved for any triads. For a triad, the four nonlinear terms appearing in the MHD equations (\ref{vel}-\ref{mag}) generate eight transfers of the type (\ref{triad_tra}), four in the energy equation and four in the cross-helicity equation, which contribute to six conserved interations, three in the energy equation and three in the cross-helicity equation. When using the Els\"asser variables, the resulting transfer terms are just the combination of the eight transfers appearing for $u$ and $b$ representation and nothing more. If we start in Els\"asser formalism and desire to recover fully the $u$ and $b$ representation, the residual energy (correlation of the co-propagating and contra-propagating Alfv\'en waves phase velocities) needs to be taken into account.

\subsection{The scale transfer functions}

To quantify the transfer between scales, we decompose the wavenumber space into a series of disjoint shells $s_\si \equiv(k_{\si-1},\ k_\si]$, similar to other works on the subject \cite{Domaradzki:1990p145, Kida:1992p958, Domaradzki:2009p557, Alexakis:2005p304, Mininni:2005p123}. The velocity field $\hat \uu^\si$ and magnetic field $\hat \bb^\si$ (note that the hat denotes Fourier transform) contained in a shell identified by the index $\si$ are found, by the use of a sharp spectral filter, to be: 
\begin{eqnarray}
\hat \uu^\si(\kk)&=\left\{ \begin{array}{rcl}
\hat \uu(\kk) & \mbox{if} &  |\kk| \in s_\si \\
0 & \mbox{if} &   |\kk| \notin s_\si
\end{array}  \right. \;, \\
\hat \bb^\si(\kk)&=\left\{ \begin{array}{rcl}
\hat \bb(\kk) & \mbox{if} &  |\kk| \in s_\si \\
0 & \mbox{if} &   |\kk| \notin s_\si
\end{array}  \right. \;.
\end{eqnarray}
The choice of a sharp filter compared to a smooth one has been investigated by Domaradzki and Carati\cite{Domaradzki:2007p134, Domaradzki:2007p133} who found that the transfer functions and energy fluxes are similar in the two cases for sufficiently compact smooth filters. The shell wavenumber boundaries are obtained from the geometrical progression: $k_\si=4 \times 2^{(\si-1)/4}$, with $k_0=0$. We obtain $N=25$ shells ($\si=1,2,...,N$) for $k_{\max}=256$ and $N=29$ for $k_{\max}=512$. The use of a geometrical progression for the shells boundaries, assures us of capturing a sufficiently localized signal in both spectral and real space. For a unit linear separation for the shell boundaries, the angle integrated quantities would be recovered. When working with angle integrated quantities, the designation band is usually employed instead of shells, as the wavenumber space decomposition is seen as selecting bands of the one-dimensional wavenumber space. The real space representation $\uu^\si(\xx)$ of the shell filtered velocity field, corresponds to a characteristic velocity for a scale $\delta k_\si$, with the shell based scales $\delta k_\si$ separated as $\delta k_{\si-1}/\delta k_\si \sim 2^{1/4}$. This separation should be considered as the smallest separation between scales that we account for and not as a scaling directly linked to turbulence itself. In real space, the total field can be recovered by summing over each shell filtered contribution:
\begin{align}
\uu(\xx)=\sum_{\si=1}^N\uu^\si(\xx)\;,\ \ \ \ \bb(\xx)=\sum_{\si=1}^N\bb^\si(\xx)\;.
\end{align}

Numerically the transfer occurring between the shell filtered fields $\hat {\bf X}^\si$, $\hat {\bf Y}^\sii$ and $\hat {\bf Z}^\siii$, is computed as,
\begin{align}
S^X_{Y,Z}(\si|\sii|\siii)&=\sum_{\kk \in s_\si}\Re\{ i\kk \cdot \widehat{ {\bf Z}^\siii {\bf Y}^\sii}(\kk) \cdot  \hat{\bf X}^{\si}(\kk)\}\;.
\label{trans3scale}
\end{align}
The triple transfer between shells $S^X_{Y,Z}(\si|\sii|\siii)$, ignoring the fields nature but taking into account their position in the interaction, has a more precise interpretation as the transfer to shell $\si$ from shell $\sii$, mediated by (through advection by) shell $\siii$. The contribution to a scale from the other two scales, regardless on the role they play in the interaction can also be defined as,
\begin{align}
& T^X_{Y,Z}(\si|\sii,\siii)=\frac{1}{2} \left[S^X_{Y,Z}(\si|\sii|\siii)+S^X_{Y,Z}(\si|\siii|\sii)\right]\ ,\label{trans_sim}
\end{align}
and can be useful when employing infinitesimally thin shells, as the recovered symmetry would allow one to work with this quantity in direct analogy to the triad transfer (\ref{triad_tra}). 

The function $S^X_{Y,Z}(\si|\sii|\siii)$ is computed numerically from solutions of the MHD equations and forms the basis of our analysis. Knowing $S^X_{Y,Z}(\si|\sii|\siii)$ allows us to compute the shell-to-shell transfers by summing over all possible advective shells.
\begin{align}
& P^X_{Y,Z}(\si|\sii)=\sum_{\siii} S^X_{Y,Z}(\si|\sii|\siii) \;.
\end{align}
\begin{figure}[tb]
\centering
\includegraphics[width = 0.48\textwidth]{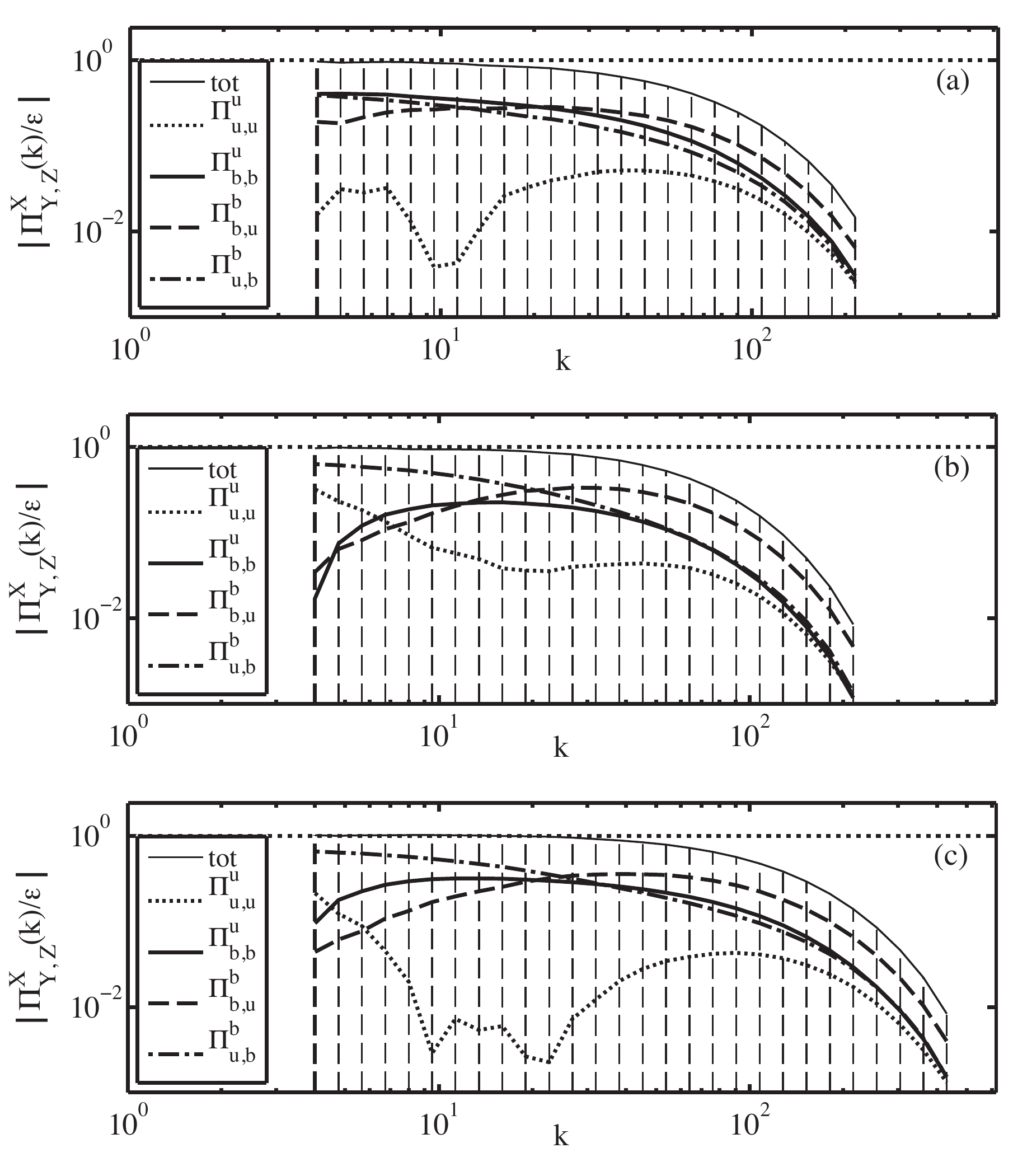}
\caption{The total flux for run I (a), run III (b) and run IV (c) and the contributions made by the four nonlinear terms, displayed in absolute values. The vertical lines depict the shell boundaries. For run I, up to shell 7 and for run IV, between shell 6 and shell 12, the flux $\Pi^u_{u,u}$ is negative.}
\label{fig011}
\end{figure}
The shell-to-shell function $P^X_{Y,Z}(\si|\sii)$ can be seen as a transfer from shell $\sii$ of field $Y$ to shell $\si$ of field $X$ and possesses the antisymmetry property $P^X_{Y,Z}(\si|\sii)=-P^Y_{X,Z}(\sii|\si)$. 

Similarly, the net transfer to a shell $T^X_{Y,Z}(\si)$ can be found by summation over $\sii$ and $\siii$,
\begin{align}
& T^X_{Y,Z}(\si)=\sum_{\sii} P^X_{Y,Z}(\si|\sii)=\sum_{\sii}\sum_{\siii} S^X_{Y,Z}(\si|\sii|\siii)\;, \label{nettrans}
\end{align}
and can be seen as the shell integrated transfer spectra. When summing the net transfer (\ref{nettrans}) over $\si$, which is equivalent to integrating the nonlinear transfer over the entire space, we obtain zero only for the interactions that are conserved in a triad. This fact requires extra care from us when defining and interpreting the energy fluxes through a shell boundary surface, as not all fluxes go to zero in the UV limit (large wavenumber limit).

The flux trough a shell boundary ($k_\kc$) is then defined by partial summing the transfer band spectra $T^X_{Y,Z}(\si)$,
\begin{align}
\Pi^X_{Y,Z}(k_\kc)=\sum_{\si=\kc+1}^{N} T^X_{Y,Z}(\si)=\sum_{\si=\kc+1}^{N} \sum_{\siii=1}^{N} \sum_{\sii=1}^{N} S^X_{Y,Z}(\si | \sii | \siii)\;.
\end{align}
These fluxes for runs I and III are shown in Fig.\ref{fig011}.

\subsection{The flux locality functions}

\begin{figure*}[tb]
\centering
\includegraphics[width = 0.85\textwidth]{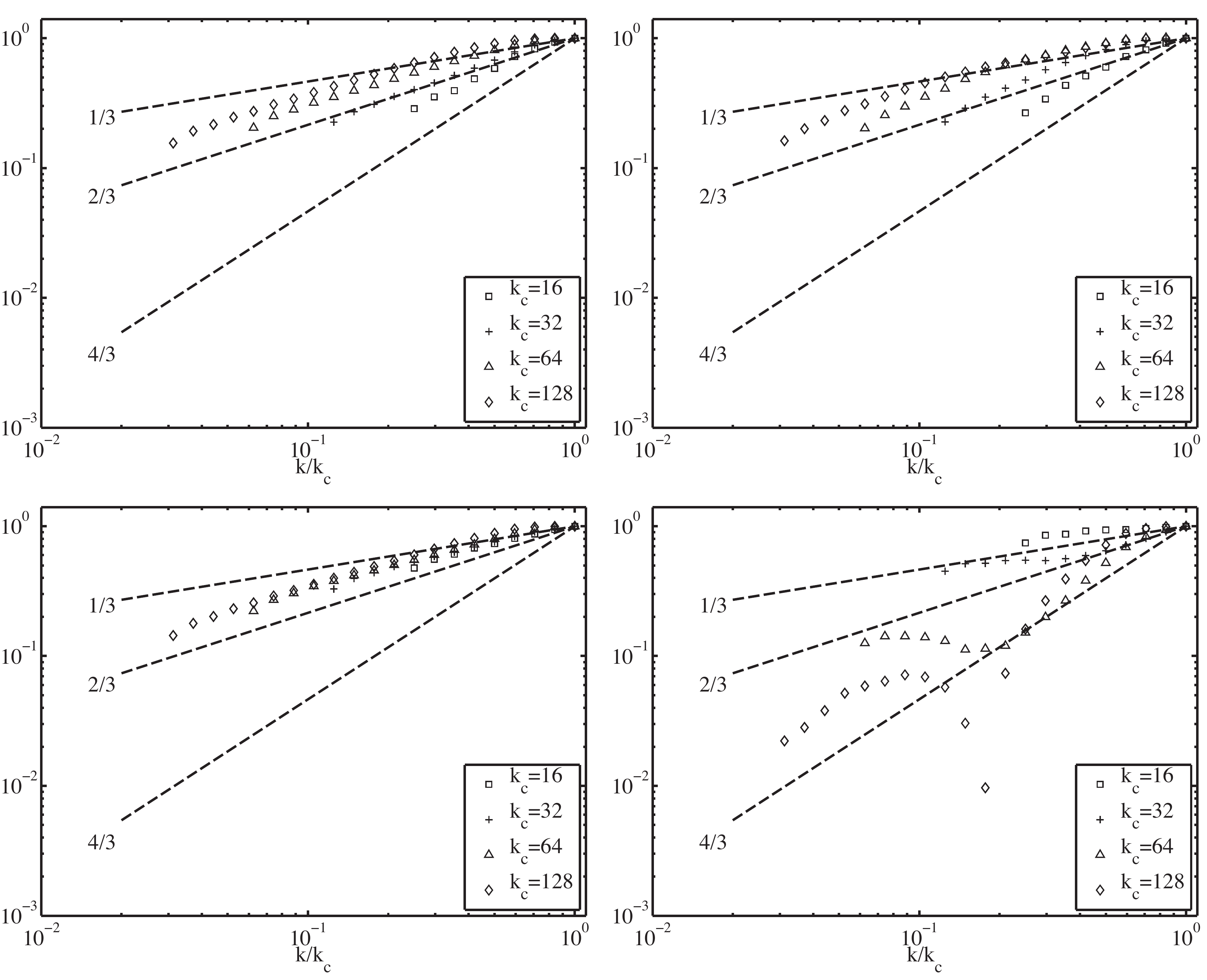}
\caption{Infrared locality functions for run III.  $\Pi_{\mbox{\scriptsize ir}}\,^u_{u,u}(k|k_c)$ (top left), $\Pi_{\mbox{\scriptsize ir}}\,^u_{b,b}(k|k_c)$ (top right),$\Pi_{\mbox{\scriptsize ir}}\,^b_{b,u}(k|k_c)$ (bottom left) and $\Pi_{\mbox{\scriptsize ir}}\,^b_{u,b}(k|k_c)$ (bottom right) are presented in absolute value and are normalized by their respective fluxes through $k_\kc$. }
\label{fig020}
\end{figure*}

\begin{figure*}[tb]
\centering
\includegraphics[width = 0.85\textwidth]{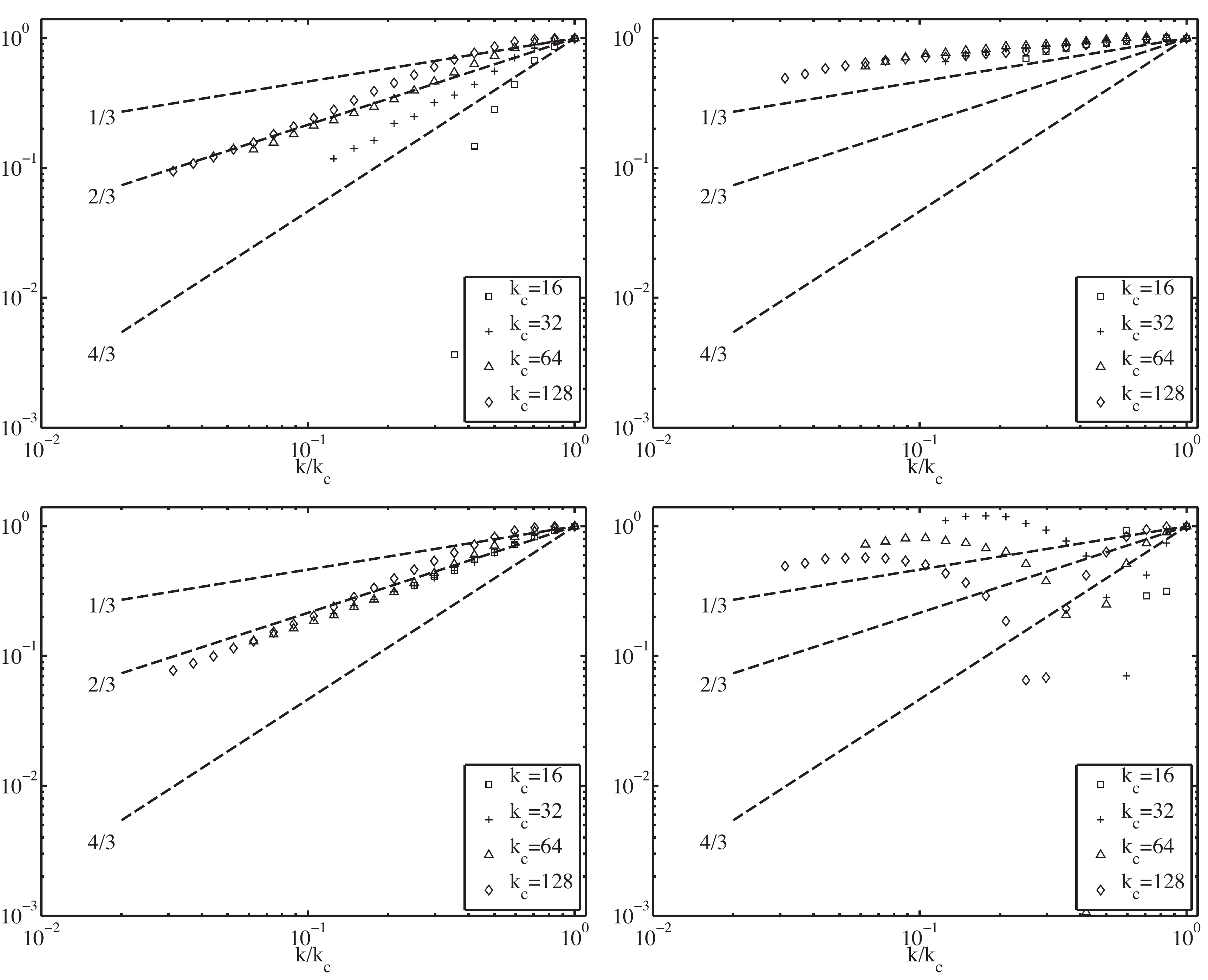}
\caption{Infrared locality functions for run I.  $\Pi_{\mbox{\scriptsize ir}}\,^u_{u,u}(k|k_c)$ (top left), $\Pi_{\mbox{\scriptsize ir}}\,^u_{b,b}(k|k_c)$ (top right),$\Pi_{\mbox{\scriptsize ir}}\,^b_{b,u}(k|k_c)$ (bottom left) and $\Pi_{\mbox{\scriptsize ir}}\,^b_{u,b}(k|k_c)$ (bottom right) are presented in absolute value and are normalized by their respective fluxes through $k_\kc$. }
\label{fig021}
\end{figure*}

\begin{figure}[tb]
\centering
\includegraphics[width = 0.4\textwidth]{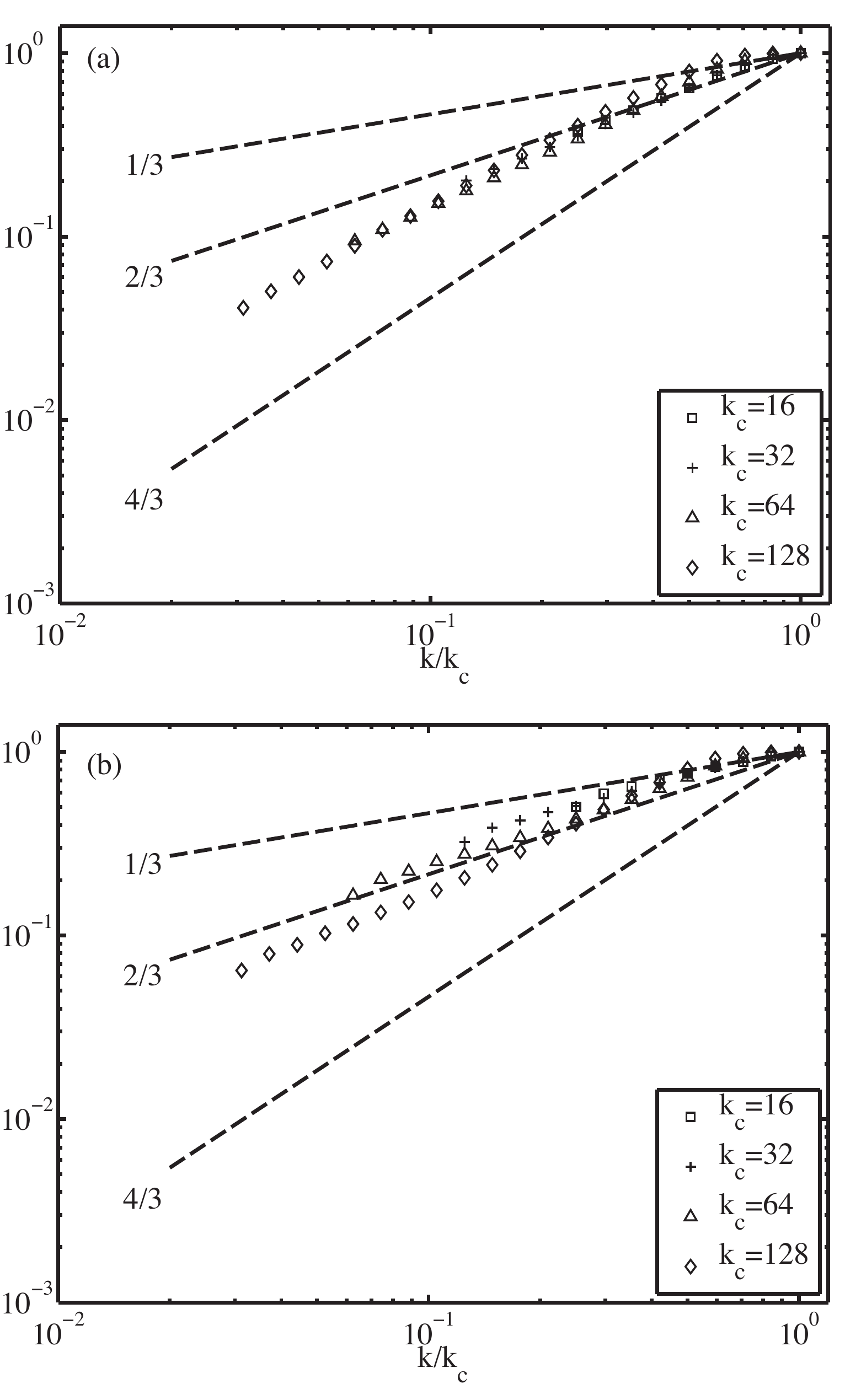}
\caption{Infrared locality functions for the total conversion term $\Pi_{\mbox{\scriptsize ir}}\,^u_{b,b}(k|k_c)+\Pi_{\mbox{\scriptsize ir}}\,^b_{u,b}(k|k_c)$; run I (a) and run III (b).}
\label{fig022}
\end{figure}

The infrared locality function is defined by taking a probe wavenumber boundary $k_\kp$, so that $k_\kp \le k_\kc$, and it measures the contribution to the flux trough $k_\kc$  from triads of modes with at least one wavenumber less than $k_\kp$,
\begin{align}
\Pi_{\mbox{\scriptsize ir}}\,^X_{Y,Z}(k_\kp|k_\kc)=\sum_{\si=\kc+1}^{N}\left[\sum_{\sii=1}^{N}\right. \sum_{\siii=1}^\kp S^X_{Y,Z}(\si | \sii| \siii) \nonumber\\
+\left.  \sum_{\sii=1}^{\kp}\sum_{\siii=\kp+1}^N S^X_{Y,Z}(\si | \sii| \siii) \right] \;.
\end{align}
In the second term, the sum over shell $\siii$ starts from $\kp+1$ to avoid double counting. For the limit $k_\kp \rightarrow k_\kc$, we recover the flux for all terms. It is customary to normalize the locality functions to the flux trough $k_c$, which from the definition shows that is one for $k_p=k_c$ and decreases for $k_p/k_c<1$,
\begin{align}
\lim_{k_p\rightarrow k_c}\left[\frac{\Pi_{\mbox{\scriptsize ir}}\,^X_{Y,Z}(k_p|k_c)}{\Pi^X_{Y,Z}(k_c)}\right]=\lim_{k_p\rightarrow k_c}\left[\frac{\Pi_{\mbox{\scriptsize ir}}\,^X_{Y,Z}(k_p|k_c)}{\Pi_{\mbox{\scriptsize ir}}\,^X_{Y,Z}(k_c|k_c)}\right]=1\;.
\end{align}
Similarly, the ultra-violet locality function $\Pi_{\mbox{\scriptsize uv}}(k_p|k_c)$, which accounts for the contributions of small scales to the transfer can be defined. The UV functions are strongly influenced by the dissipation (or more correctly by the shape of the flux which naturally decreases for high $k$ in dissipative systems) and do not exhibit a clear scaling. From our numeric analysis, a lower limit of $1/3$ can be safely inferred for the UV locality exponent, without excluding the possibility that the actual asymptotic values are closer to $2/3$. The same dependency of the scaling slope clarity on the shape of the fluxes is experienced by the IR locality function, as well. However, due to the forcing mechanism used, these effects are found to influence the IR locality exponent scaling to a lesser extent and for this reason we focus in this work on the analysis of the IR locality exponents.

For the kinetic forced run, the IR locality functions are shown for the four energy fluxes in Fig.\ref{fig020}. For the flux of kinetic energy ($\Pi^u_{u,u}$) and the flux of magnetic energy ($\Pi^b_{b,u}$) a $2/3$ scaling can be observed. The conversion terms, \ie kinetic to magnetic ($\Pi^u_{b,b}$) and magnetic to kinetic ($\Pi^b_{u,b}$) fluxes, exhibit a more complex picture. For these two terms the locality scaling seems to be between $1/3$ and $4/3$. 
Looking now at the kinetic-inductive forced run, Fig.\ref{fig021}, we see that, while the kinetic and magnetic energy fluxes asymptote to the same $2/3$ value, the conversion terms show a clear $1/3$ scaling. Since only absolute values are plotted in these figures, the presence of a cusp is indicative of a change in the sign of the locality functions. As the locality functions are normalized by their respective flux through $k_\kc$, the change in sign indicates a change in the nature of the contributions to the flux across the cutoff scale.

To obtain a better physical understanding of the $1/3$ scaling value, we plot in Fig.\ref{fig022} the IR locality scaling for the entire conversion term (containing the interactions of both of the two cross-field terms). We remind the reader that only when considered in its totality does the cross-field interaction is energy conserving. Because of this conservation and cancelation effects, the global scaling for the conversion flux drops to values closer to $2/3$ or even larger denoting a more local behavior.

\section{A detailed analysis of at the total flux}

The cancelation nature of the fluxes and the effect this has on the locality properties require an additional study. To maximize the lesson learned from HD turbulence and use the intuition gained by the community in this field, we look at the total flux. For MHD, we define the total net energy transfer to a shell as:
\begin{align}
& T(\si)=T^u_{u,u}(\si)+T^b_{b,u}(\si)-T^u_{b,b}(\si)-T^b_{u,b}(\si) \;. \label{totnettrans}
\end{align}
For MHD turbulence, the total net transfer $T(\si)$ and the afferent quantities (\eg total triple transfer between shells, total shell-to-shell transfer) are the ones that have a behavior similar to the respective quantities found for HD turbulence. For each of the quantities introduced in the previous section, a similar sum can be performed. Alternatively, one could just define the total energy triad transfer as the sum of the four energy triad transfers possible and build up the framework for the total energy exchange among scales. From this point on, quantities for which we omit the ${X,Y,Z}$ labels refer to total quantities. 

To better understand the contribution to a flux trough a shell boundary ($k_\kc$), we look at the total flux, defined as: 
\begin{align}
\Pi(k_\kc)=\sum_{\si=\kc+1}^{N} T(\si)=\sum_{\si=\kc+1}^{N} \sum_{\sii=1}^{N} \sum_{\siii=1}^{N} S(\si | \sii | \siii)\;.
\end{align}
For the three runs, the total fluxes are shown in Fig.~\ref{fig011}.
Writing in detail the second two sums in regard to the shell index $\kc$ we find four terms, each one with a different physical interpretation,
\begin{align}
\sum_{\sii=1}^{N} \sum_{\siii=1}^{N} S(\si | \sii| \siii)&=\sum_{\sii=1}^\kc \sum_{\siii=1}^\kc S(\si | \sii| \siii) \ \ \ \ \ \ \ \ \ \ \Big{\}}(i) \nonumber \\
&+\sum_{\sii=1}^\kc \sum_{\siii=\kc+1}^{N} S(\si | \sii| \siii)  \ \ \ \ \ \ \, \Big{\}}(ii) \nonumber\\
&+\sum_{\sii=\kc+1}^{N} \sum_{\siii=1}^\kc S(\si | \sii| \siii) \ \ \ \ \ \ \  \Big{\}}(iii)  \nonumber\\
&+\sum_{\sii=\kc+1}^{N} \sum_{\siii=\kc+1}^{N} S(\si | \sii| \siii) \ \ \ \Big{\}}(iv) \;,
\end{align}
The first term $(i)$ contributes to the flux as the transfer from large scales to small scales, advected by the large scales, while the second term $(ii)$ represents the transfer from large scales to small scales, advected by the small scales. The third term $(iii)$ can be seen as the transfer caused by the large scales advection of the small scales and in general is found to be close to zero. The last term $(iv)$, which could be seen as the transfer from small scales to small scales by the advection of the small scales, is exactly zero due to the conservation of the total energy for any closed set of modes interacting only among themselves. The contributions to the total flux can be seen in Fig.\ref{fig001}. 

\begin{figure}[b]
\centering
\includegraphics[width = 0.48\textwidth]{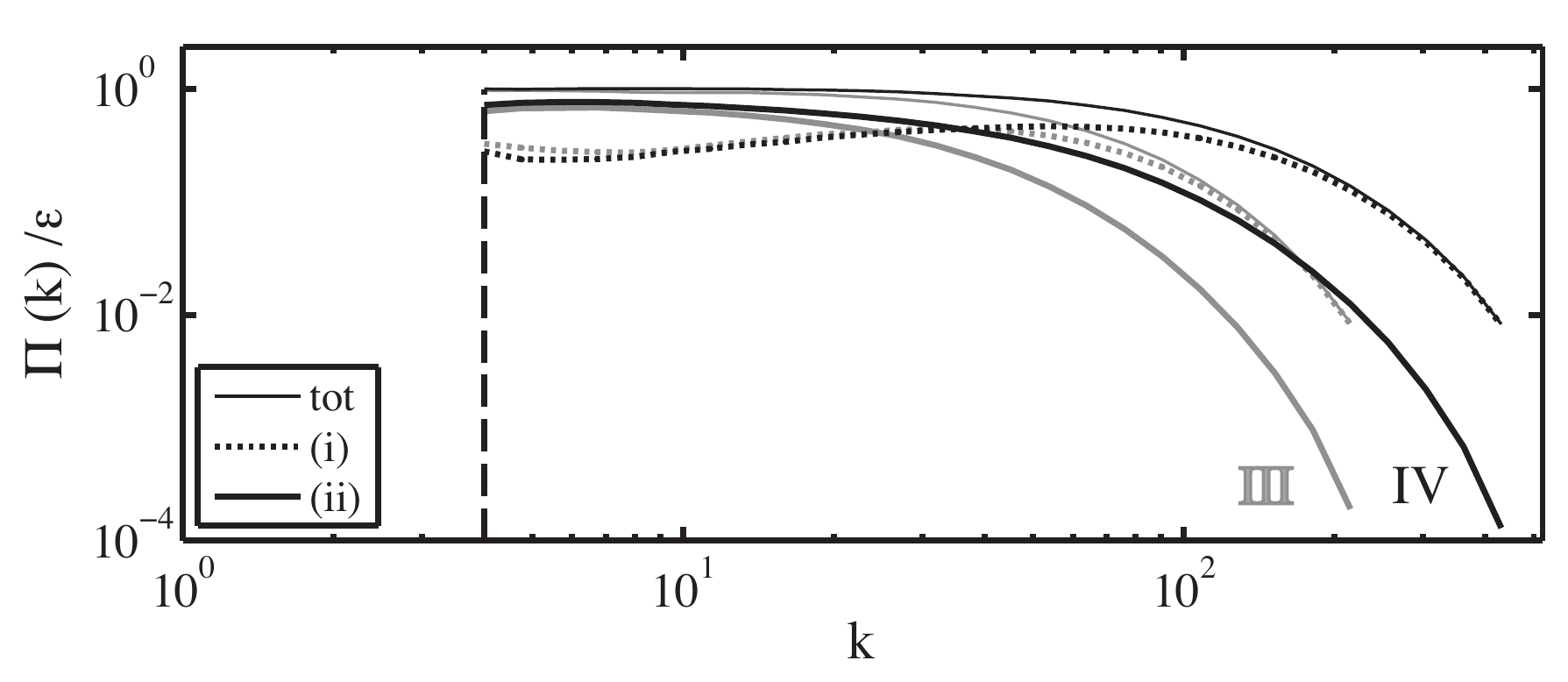}
\caption{Total flux for run III (gray) and run IV, displaying the contributions made by the main terms. The $(iii)$ terms are $10^{-6}$ times smaller compared to the other two nonzero terms and fluctuate around zero.}
\label{fig001}
\end{figure}

To account for the degree of locality of the total flux, we look at the locality properties of the two terms (fluxes) that bring the main contribution $(i)$ and $(ii)$, 
\begin{align}
\Pi(k_\kc)&\approx \Pi^{(i)}(k_\kc)+\Pi^{(ii)}(k_\kc)\nonumber \;,\\
\Pi^{(i)}(k_\kc)&=\sum_{\si=\kc+1}^{N} \sum_{\sii=1}^\kc \sum_{\siii=1}^\kc S(\si | \sii| \siii) \;,\nonumber\\
\Pi^{(ii)}(k_\kc)&=\sum_{\si=\kc+1}^{N} \sum_{\sii=1}^\kc \sum_{\siii=\kc+1}^{N} S(\si | \sii| \siii) \;.
\end{align}
Knowing the value of the flux through a shell surface $k_\kc$ we want to know how much of this flux is due to modes with wavenumbers close but smaller then $k_\kc$. For this purpose we take a probe (test) wavenumber boundary $k_\kp$, so that $k_\kp \le k_\kc$, and we measure the contribution to the flux trough $k_\kc$ from modes with wavenumber less than $k_\kp$, similar to the philosophy of the IR locality functions. By keeping $k_\kc$ fixed and varying $k_\kp$, we should obtain a smaller and smaller contributions. The rate at which the contributions become smaller is related to the locality of the flux.

\begin{figure}[tb]
\centering
\includegraphics[width = 0.4\textwidth]{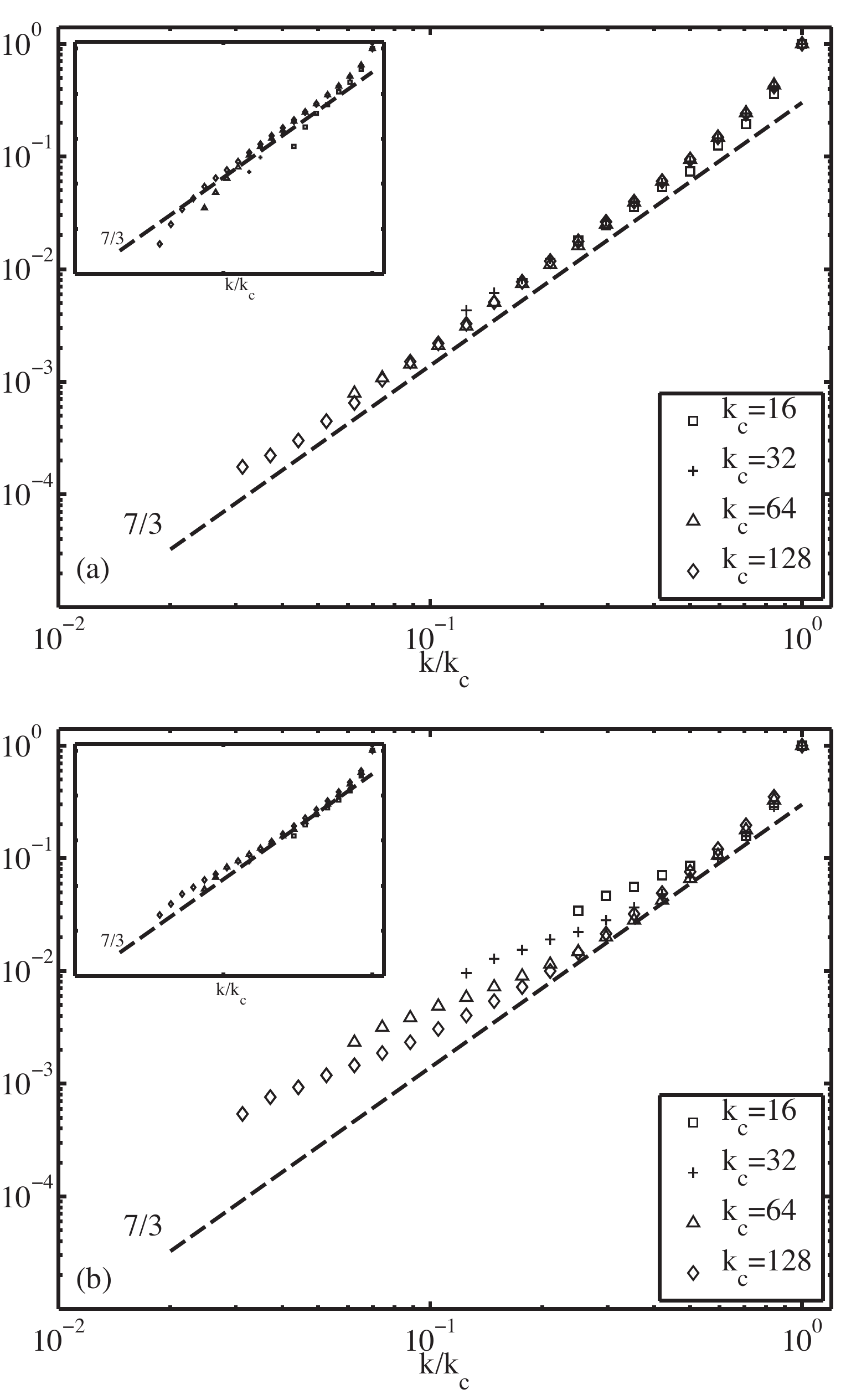}
\caption{Locality scaling for the contribution of energy across the $k_c$, advected by the entire large scale region ($\Pi^{(i)}_\sii(k_\kp|k_\kc)$): run I (a) and run III (b). Inset pictures show the same pictures without the contribution of the first shell responsible for the forcing.}
\label{fig003}
\end{figure}

In the case of $\Pi^{(i)}$ we can take the probe on the giver shell index $\sii$,
\begin{align}
\Pi^{(i)}_\sii(k_\kp|k_\kc)=\sum_{\si=\kc+1}^{N} \sum_{\sii=1}^\kp \sum_{\siii=1}^\kc S(\si | \sii| \siii) \;,
\end{align}
 or on the advecting shell index $\siii$
\begin{align}
\Pi^{(i)}_\siii(k_\kp|k_\kc)=\sum_{\si=\kc+1}^{N} \sum_{\sii=1}^\kc \sum_{\siii=1}^\kp S(\si | \sii| \siii) \;.
\end{align}
%
\begin{figure*}[t]
\centering
\includegraphics[width = 0.85\textwidth]{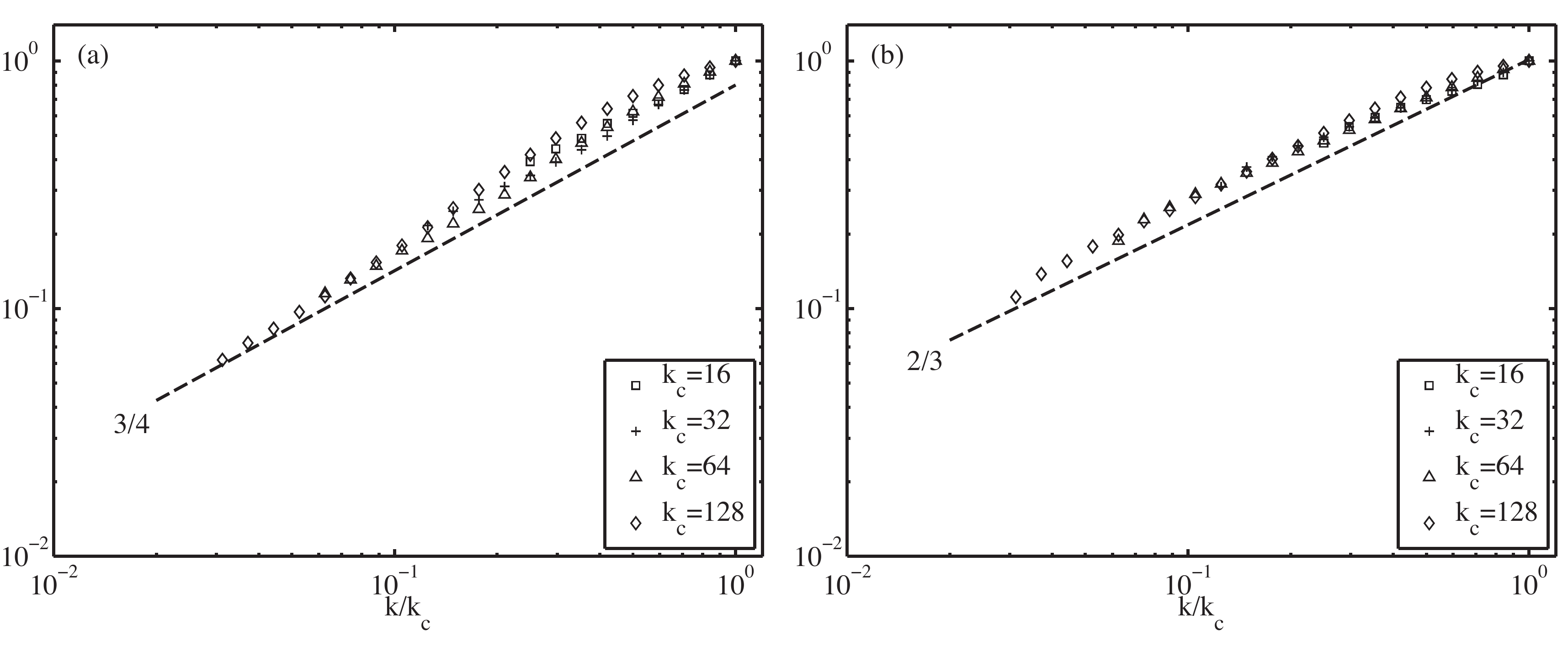}
\caption{Locality scaling for the contribution of energy advected across the $k_c$, given by the entire large scale region ($\Pi^{(i)}_\siii(k_\kp|k_\kc)$): run I (a) and run III (b).}
\label{fig004}
\end{figure*}
The locality nature of the partial flux $\Pi^{(i)}$ is related to both effects, however, performing the probe variation separately allows us to understand the locality nature of these two effects individually. First, by varying the probe on the giver shell and keeping the advection to the entire large scales region, we transfer energy from increasingly separated scales. From Fig.\ref{fig003}, we see that the locality function scales as $7/3$, which would indicate a highly local behavior. Second, we vary the probe on the advecting shell while keeping the entire large scale interval as the giver of energy. In Fig.\ref{fig004} we see that the advection brings a more nonlocal behavior to the partial flux. When considered together it is the more nonlocal behavior that dominates the asymptotic behavior.

For the partial flux $\Pi^{(ii)}$, a more pronounce nonlocal behavior has been found. The scaling is close to a $2/3$ value for the kinetic forcing and a $3/4$ value for the kinetic-inductive forcing. The value of $3/4$ for the IR locality exponent is consistent with the theoretical value found by Aluie and Eyink\cite{Aluie:2010p946} for fields that scale as $k^{-1/4}$. 

From the locality picture of the partial fluxes we see that the advecting effects are the ones responsible with the value of the IR scaling exponent for the total flux. Combining all these behaviors (advecting and advected) results in a scaling close to $2/3$ obtained for Kraichnan IR locality functions for the total energy flux of MHD turbulence.

\section{Conclusions and discussion}

After investigating the locality properties of MHD energy fluxes, it is found that asymptotically the dynamics tend to be dominated by local interactions. The nonlocal interactions that apparently exist, cancel themselves out. However, the locality is much weaker compared to the case of HD turbulence, which is characterized by the scaling exponent of $4/3$. When using a velocity proportional force, two distinct exponents are observed for MHD turbulence, $1/3$ and $2/3$ for various fluxes. The $1/3$ exponent is even more obvious for the kinetic-inductive forcing mechanism. 

To better understand these values we investigated the total energy flux. The lesson taught by the analysis of the total flux showed us that the locality of a flux can be seen as the locality of the contribution of each effect that makes up that flux. For similar level contributions, it is the most non-local channel in a flux that imprints the overall locality behavior. For MHD turbulence, the overall locality seams to be close to $2/3$ and confirms the analysis done by Aluie and Eyink\cite{Aluie:2010p946} who showed a $2/3$ scaling behavior for the individual interactions of HD and MHD turbulence. Since for MHD turbulence, the interplay between velocity and magnetic field gives rise to much smoother correlated fields, the decorrelation effects due to the averaging procedure are less pronounced, which results in a smaller global locality exponent and justify the $2/3$ value found for the various energy fluxes compared to the $4/3$ value found for the energy flux of HD turbulence. This can be seen best by looking at the total flux, the corresponding flux for the velocity flux in hydrodynamical case. Various non-locality contributions and force influence cancel themselves, making the total flux scaling exponent the most reliable observation. 

The observed $1/3$ value was initially thought to be caused by the forcing. Assuming that the $1/3$ locality exponent is indeed due to the external force or due to the pollution of the inertial range by the large scale strains and that for a proper inertial-inductive range the $2/3$ exponent index is recovered, this still represents a major problem for adequate modeling. A proper inertial-inductive range is so slow to appear that a huge number of modes need to be solved for any LES type model. For practical reasons, if the forcing range pollutes the inertial-inductive range to this extent it is better to take into account the more pronounced nonlocal behavior of the individual fluxes rather then trying to reach the inertial-inductive range. We believe that the $1/3$ scaling exponents is found due to the consideration of the conversion terms individually and is not just an artifact of the force. When triads are summed in a non-conservative way, fluxes of energy naturally appear. This fact may represent another problem for practical modelling. In MHD shell models\cite{Lessinnes:2009p805,Lessinnes:2009p719} the two conversion terms are still accounted for separately (a stronger nonlocal character) but their interaction is still considered to be mainly with only the closest neighbor shells, in the same way as in the case of the more local kinetic and magnetic terms. Allowing the cross-field terms to couple to more distant shells might be more appropriate as the level of contribution neglected to a flux would be the same for all individual terms.

A question remains: Is there any physical significance for the $1/3$ exponent seen for the energy conversion terms? If we try to interpret MHD turbulence as scattering of contra-propagating Alfv\'en waves, then we actually work with the wave phase velocities regardless of actually employing Els\"asser formalism or not (which just acknowledges the fact that we are mixing $u$ and $b$ in the definition of the wave phase velocities). For this interpretation there is no $1/3$ scaling. Moreover, in this interpretation all the four energy fluxes and any combination of them are just partial fluxes, pieces that make up the entire physical picture. Only the energy of a wave should be considered in this case and that energy is the Elsasser pseudo-energy (which is equal to the total energy only in the absence of cross-helicity) and for which the respective fluxes have a scaling exponent close to $2/3$. Complementary, if we just look at the interactions in a triad and assign a physical interpretation to each conserved interaction (\eg kinetic energy exchange, magnetic energy exchange, energy conversion), we need to consider the two conversion terms together and we again do not obtain the $1/3$ scaling since the nonlocal contribution cancel themselves. 

However, when MHD equations for $u$ and $b$ are used, as is the case most often, the scaling for individual terms appears to be a legitimate question. Looking at the two energy evolution equations separately (notwithstanding the arguments that the two are not strictly redistributive by themselves) we can state that the kinetic energy of the system is modified by a flux (generated by the the entire nonlinear rhs) which possesses a $1/3$ locality scaling exponent, since the $u$ to $b$ flux is larger in magnitude and has the strongest nonlocal behavior. This information may not be adequate for the physical interpretation of the energy transfers, but it may be useful for modeling if $u$ and $b$ equations are used and one tries to model effects of each term separately.

\acknowledgments{
This work has been supported by the contract of association EURATOM-Belgian State. The content of  the publication is the sole responsibility of the authors and it does not necessarily  represent the views of the Commission or its services. BT would like to acknowledge Universit\'e Libre de Bruxelles for being the hosting institution during the preparation of this work.


%

\end{document}